\documentclass[12pt]{article}

\def\slash#1{\setbox0=\hbox{$#1$}#1\hskip-\wd0\hbox to\wd0{\hss\sl/\/\hss}}
\usepackage[all]{xy}
\usepackage{verbatim}
\usepackage[pdftex]{color, graphicx}
\usepackage{amsfonts}
\usepackage{latexsym}
\usepackage{amsmath}
\usepackage{amssymb}
\usepackage{cite}

\usepackage{epsf, cite}

\makeatletter
\renewcommand\section{\@startsection {section}{1}{\z@}%
                                   {-3.5ex \@plus -1ex \@minus -.2ex}
                                   {2.3ex \@plus.2ex}%
                                   {\normalfont\large\bfseries}}
\renewcommand\subsection{\@startsection{subsection}{2}{\z@}%
                                     {-3.25ex\@plus -1ex \@minus -.2ex}%
                                     {1.5ex \@plus .2ex}%
                                     {\normalfont\bfseries}}
\makeatother

\def\lbldef#1#2{\expandafter\gdef\csname #1\endcsname {#2}}

\def\href#1#2{#2}

\newcommand\be{\begin{equation}}
\newcommand\ee{\end{equation}}
\newcommand{\bea}{\begin{eqnarray}}
\newcommand{\eea}{\end{eqnarray}}

\newcommand{\p}{\partial}

\def\id{\protect{{1 \kern-.28em {\rm l}}}}

\def\Tr{{\textrm{Tr}} }

\def\unit{\relax{\rm 1\kern-.26em I}}

\def\Z {\bb{Z}}

\def\id{\protect{{1 \kern-.28em {\rm l}}}}

\def\a{\alpha}
\def\b{\beta}
\def\c{\gamma}
\def\d{\delta}
\def\e {\epsilon}

\def\r{\rho}
\def\t{\tau}
\def\th{\theta}
\begin{document}
\pagestyle{plain}
\begin{titlepage}

\begin{center}

\hfill{QMUL-PH-2009-07} \\

\vskip 1cm

{{\Large \bf Membranes with a Boundary}} \\

\vskip 1.25cm {David S. Berman\footnote{email: D.S.Berman@qmul.ac.uk},
 and
Daniel C. Thompson\footnote{email: D.C.Thompson@qmul.ac.uk}}
\\
{\vskip 0.2cm
Queen Mary, University of London,\\
Department of Physics,\\
Mile End Road,\\
London, E1 4NS, England\\
}
\end{center}
\vskip 1 cm

\begin{abstract}
\baselineskip=18pt\
We investigate the recently developed theory of multiple membranes.  In
particular, we consider open membranes, i.e. the theory defined on a
membrane world volume with a boundary.

We first restrict our attention to the gauge sector of the theory.  We
obtain a boundary action from the Chern-Simons terms.

Secondly, we consider the addition of certain boundary terms to
various Chern-Simons theories coupled to matter.
These terms ensure the full bulk plus boundary action has the correct
amount of supersymmetry.
For the ABJM model, this construction motivates the inclusion of a
boundary quartic scalar potential.  The boundary dynamics obtained
from our modified theory produce Basu-Harvey type equations describing
membranes ending on a fivebrane.

The ultimate goal of this work is to throw light on the theory of fivebranes
using the theory of open membranes.

\end{abstract}

\end{titlepage}

\pagestyle{plain}

\baselineskip=19pt
\tableofcontents

\section{Introduction}
For over a decade since the discovery of M-theory, the search for an
adequate description of multiple and coincident membranes, was fruitless. For a review of branes in M-theory see \cite{Berman:2007bv}. Recently, a significant breakthrough in this direction was made by Bagger and Lambert (BL) \cite{Bagger:2006sk,Bagger:2007jr,Bagger:2007vi}, and separately by Gustavsson \cite{Gustavsson:2007vu},  with the remarkable discovery of a 3-dimensional maximally supersymmetric (${\cal N} = 8 $) theory endowed with $SO(8)$ R-symmetry. The BL theory is based on a novel algebraic structure called a three-algebra.  However, the conditions placed on the structure constants of this algebra, namely complete antisymmetry and a generalization of the Jacobi identity, prove to be very restrictive \cite{Gauntlett:2008uf}. Essentially, there is a unique finite-dimensional three-algebra with a positive definite metric.  With this choice of algebra, the BL can also be described in a more conventional way as supersymmetric Chern-Simons gauge theory with a $SU(2)\times SU(2)$ product gauge group and bi-fundamental matter \cite{VanRaamsdonk:2008ft}.
   This theory is thought to describe two membranes in a non-trivial orbifold-like background \cite{Lambert:2008et, Distler:2008mk}.

Much work has gone into generalising the BL theory to account for an
arbitrary number of membranes.  One approach is to relax the
positivity condition on the three-algebra.  This leads to the
so-called Lorentzian BL models \cite{Benvenuti:2008bt, Bandres:2008kj,
  Gomis:2008uv} .   Another approach, which has received much
attention, is the  \textquoteleft ABJM\textquoteright\, model
\cite{Aharony:2008ug}.  This shares many features with BL theory and,
indeed, can be built from a three-algebra by relaxing the condition of
complete antisymmetry on the structure constants \cite{Bagger:2008se}.
Like the BL theory this is a super Chern-Simons theory with
bi-fundamental matter but with a more general $G=U(N)\times U(N)$
product gauge group.  Unlike the BL theory, the only R-symmetry
manifestly is $SU(4)$ and therefore the theory only possesses ${\cal
  N} = 6 $ supersymmetry.   The theory is characterized by an integer
parameter, the Chern-Simons level $k$.   At a general level $k$ this
theory is thought to describe membranes whose transverse space is an
orbifold $\mathbb{C}^4/\mathbb{Z}_k$ and is conjectured to be the CFT
dual to  $AdS_4 \times S_7/\mathbb{Z}_k$ geometry. In the case of
$k=1$ it is proposed that the supersymmetry is enhanced to ${\cal N} =
8 $ and that the theory describes  $N$ membranes in flat space with a
geometric dual of $AdS_4 \times S_7$.  Interestingly, there is also a
't Hooft limit given by taking both $N$ and $k$ large with $\lambda= N/k$ fixed. In this limit the theory admits a dual geometric description of $AdS_4 \times \mathbb{CP}_3$.

Alongside the M2 stands another extended object of M-theory; the M5-brane.  The world volume theory of even a single M5 brane is a complicated affair.  A key reason for this complexity is the self dual two-form contained in the (2,0) tensor multiplet describing the world volume fields.  Generalizing this to any non-abelian version is harder still and it seems highly unlikely that there is a straightforward way to do so.  A full theory of multiple M5-branes currently remains a distant prospect.

An important tool in the study of M5 branes is the membrane. Open membranes end on M5 branes much as open strings end on D-branes \cite{Townsend:1995af,Strominger:1995ac}.  By studying open membranes one could hope to learn about the M5 brane.  A striking example of this idea is the derivation of the M5 brane equations of motion from demanding $\kappa$-symmetry in the open membrane action \cite{Chu:1997iw}. In the context of BL theory there has already been some interesting work in attempting to extract information about the M5 brane \cite{Ho:2008nn}.

The system of a membrane ending on a fivebrane has two alternative view points.
From the membrane perspective it is conjectured, in analogy to the
D1-D3 system in string theory, that the ending can be described as a
fuzzy $S^3$ funnel solution of the Basu-Harvey equation
\cite{Basu:2004ed,Berman:2005re} and as a solution of the mass
deformed ABJM model \cite{Gomis:2008vc}.   From the perspective of the M5, the M2 brane appears as a string like soliton of the non-linear world volume equations of motion.  This soliton is known as the self dual string (SDS) \cite{Howe:1997ue}.

This paper will work towards a further description of the
open membrane. We  will attempt to generalize the recently developed
theory of interacting membranes to the describe multiple membranes whose world volume possess a boundary.  This boundary should correspond to self dual strings.

In section 2,  we shall restrict our attention to the gauge sector of ABJM theory with a boundary.   We find that this sector can be expressed as a non-chiral WZW-type model with a central charge that scales like $N$.  This result is consistent with the expected scaling of the degrees of freedom of the self dual string.

In the remainder of the paper we consider the full ABJM theory with a
boundary.  The presence of a boundary should break half the
supersymmetry; the boundary obviously breaks translational symmetry and, since supersymmetry closes on translations, it is
inevitable that the presence of boundary will also break supersymmetry. A few moments of contemplation will reveal
that only a half the supersymmetry can be preserved.   We employ and
extend the method of \cite{Belyaev:2008xk} to supplement the ABJM
theory with an appropriate boundary action so that the correct amount
of symmetry is preserved.   We will find the BPS Basu-Harvey type equations are rediscovered as boundary conditions for the open membranes.  (This is a fairly involved procedure and before tackling ABJM we address some less supersymmetric Chern-Simons matter theories.)

In section 3, we outline the procedure of \cite{Belyaev:2008xk} which allows for the construction of 1/2 supersymmetric actions for manifolds with a boundary.  In section 4, we recap how this procedure applies to the ${\cal N}=1$ abelian Chern-Simons theory and pay special attention to boundary conditions. In section 5, we consider the ${\cal N}=1$ Chern-Simons matter theory.
In sections 6,7,8, we generalize this construction to ${\cal N}=2$ superspace and apply this to abelian Chern-Simons matter theory.
In section 9, we apply these considerations to the ABJM model
formulated in ${\cal N}=2$  superspace \cite{Benna:2008zy}.  In this
${\cal N}=2$ description we are readily able to lift the construction
of ${\cal N}=1$ case and draw some interesting conclusions.  A
downside of working with ${\cal N}=2$  is that we will only ever have
partial knowledge of the consequence of a boundary since not all the supersymmetry is manifest.   There have been
some formulations of both Bagger-Lambert and ABJM using more extended
superspaces \cite{Cherkis:2008ha,Buchbinder:2008vi,Cederwall:2008vd,Cederwall:2008xu}
and it would be interesting to address the issue of a boundary in
these formalisms. Although we do not tackle the full non-abelian gauge
sector in this paper we are able to motivate the inclusion of a
certain boundary potential for the matter fields. This paper should be
viewed as the first step in a complete rigorous study of membrane
boundary theories whose eventual goal is to learn about the fivebrane.

\section{Boundary Theory of The Gauge Sector}

As a warm up to the ideas involved we examine the pure gauge
sector. The Chern-Simons action is not gauge invariant in the presence
of a boundary unless specific boundary conditions are added that
ultimately induce physical degrees of freedom on the boundary. This is an
example of the
generic idea we wish to explore. We will produce a boundary action that allows the
preservation of the right amount of symmetries (supersymmetry) but at
first we will explore the idea with the well known case of
Chern-Simons theory.

The Chern-Simons action for a Lie algebra valued gauge field, $A$, is given by
\be
\label{CSaction}
S[A] = \frac{k}{4\pi} \int_{M} \Tr \left( A\wedge  d A + \frac{2}{3} A\wedge A \wedge A\right) \, ,
\ee
where $M$ is a 3-dimensional manifold. Demanding invariance of the path under large gauge transformations requires the Chern-Simons level $k$ be quantized.  When the theory has no boundary it is purely topological; the metric does not enter into the definition of the theory\footnote{though at a quantum level it is more subtle to show the topological nature}.  If $M = \Sigma \times\mathbb{R}$ and thus $\p M = \p \Sigma \times \mathbb{R}$, one can show that $S[A]$ depends only on the values of the gauge field on the boundary and can be expressed as a WZW model \cite{Witten:1988hf, Elitzur:1989nr, Dunne:1998qy}.  Consider the simplest case where $\Sigma$ is a disc with coordinates $(r ,\theta)$ and the $\mathbb{R}$ direction is identified with time.  In this case the boundary of $M$ is a cylinder.  After choosing $A_0=0$ as an Euler-Lagrange boundary condition one finds that $A_0$ is a Lagrange multiplier enforcing a Gauss' law constraint on the remaining components of the gauge field.  Solving the constraint and a little algebra yields the WZW model,
\be
S_{wzw}[U]  =  \frac{k}{4\pi} \int_{\p M} d\th d\t \textrm{Tr} ( U^{-1} \p_\th U U^{-1} \p_t U )+  \frac{k}{12\pi} \int_{M} \textrm{Tr} ( U^{-1} d U )^3 \, ,
\ee
 where $U$ is group valued.  The second term at first seems to suggest that the theory depends on the value of $U$ across the whole of $M$. This is illusory; under a change of extension the action differs only by an integer multiple of $2\pi$ and leaves a path integral unaltered \cite{Witten:1983ar}.
An important observation is that this is a chiral WZW model. The kinetic term is non standard since it is first order in time derivatives. Careful consideration of the symmetries of this theory shows that only a left action of the group given by $U(\th, t) \rightarrow V(\th)U(\th, t)$ is a true global symmetry and this gives rise to a chiral current algebra \cite{Elitzur:1989nr}. We note however, that although the kinetic term is non-standard, we do not obviously have the equation of motion for a chiral boson \cite{Sonnenschein:1988ug}. We will return to this issue presently.

The gauge kinetic term of the ABJM (and BL) theory is similar to the above Chern-Simons action.  However, the gauge group is now a product $G=U(N)\times U(N)$, and there are two gauge fields, $A$ and $\hat{A}$; one for each factor.  The kinetic term is now
\be
S = \frac{k}{4\pi} \int_{M} \Tr \left( A\wedge  d A + \frac{2}{3} A\wedge A \wedge A  \right)-  \frac{k}{4\pi} \int_{M} \Tr  \left(\hat A\wedge  d \hat A + \frac{2}{3} \hat A\wedge \hat A \wedge\hat A  \right)\, ,
\ee
with the essential difference in sign between the two factors.   The case of BL $SU(2) \times SU(2)$ theory was discussed in \cite{Berman:2008be} where it was proposed that the associated WZW model has a six-dimensional target space. This target space could perhaps be related to the M5 world volume geometry back reacted by multiple self-dual strings.  That analysis was, of course, predicated on the conjecture that the  $SU(2) \times SU(2)$  theory of the BL was unique and could describe multiple M2s.  The subsequent emergence of the more general ABJM model, and the understanding that BL describes only two membranes, invites us to revise such an interpretation.

A key observation to make is that the Chern-Simons action (\ref{CSaction}) is not parity invariant.  We will denote our coordinates in three dimensions as $x^\mu = \{x^0, x^1 , x^3 \}$. The action of parity is a refelection in either spatial coordinate and is given by\cite{Dunne:1998qy}
\bea
{\cal P}: x= \{x^0, x^1 , x^3 \}  &\mapsto& x^\prime = \{x^0, x^1, -x^3\} \, ,\\
{\cal P}: A = \{A_0(x), A_1(x) , A_3(x)\}   &\mapsto& A^\prime = \{ A_0(x^\prime), A_1(x^\prime) \, ,  -A_3(x^\prime) \}\, \, .
\eea
On spinors this acts with the multiplication by $\c^3$.  Under this action the Chern-Simons term picks up a minus sign.  The  ABJM theory is then parity invariant with the additional identification that parity also swaps the two gauge fields \cite{VanRaamsdonk:2008ft}.

When we have a boundary, which we will henceforth assume to be in the
$x^3$ direction, the Lorentz symmetry is broken and it makes a
difference which spatial coordinate one chooses to reflect in.
If, instead of the above parity operation (which we shall denote as
${\cal P}^{(3)}$)
we choose to reflect the $x^1$ direction with a corresponding operator
${\cal P}^{(1)}$
then we can develop a notion of chirality.
We may define chirality projectors on spinors as $P_\pm = \frac{1}{2}
(1 +  \c^3)$.
Then ${\cal P}^{(1)}$ acts essentially by swapping plus with minus
i.e. it switches chirality and swaps over light cone coordinates $x_\pm = x_0 \pm x_1$. For example on a spinor,  ${\cal P}^{(1)}
:\psi_+ = P_+\psi \mapsto P_+\gamma^1 \psi =  \gamma^1P_-\psi =
\gamma^1\psi_-$.  Again for the ABJM model these actions are to be
combined with a switch of fields.

With this in mind we understand that the $G=U(N)\times U(N)$ gauge
kinetic terms of the  ABJM model produces
two chiral WZW models.  The three dimensional parity transformation
descends to a
two dimensional chirality transformation and has the effect of
switching the two WZW factors.
In short, the two gauge fields and relative sign of the levels are exactly what is needed to construct a full non-chiral string. A useful observation at this stage is that in deriving the WZW model we choose a boundary condition $A_0 = \hat{A_0} = 0$. This boundary condition is compatible with parity invariance and as a result the final boundary theory is non-chiral.

It is interesting to note that one would arrive at a similar non-chiral string by considering a single gauge group but when  $\Sigma$ has the topology of an annulus rather than a disc \cite{Elitzur:1989nr}.

The fact that the left and right moving sectors are decoupled means that the central charge is given
by the usual result (see e.g. \cite{DeVecchia})
\be
c = \frac{k  \textrm{dim} SU(N) }{ k + \tilde{h}_{SU(N)}} = \frac{ k( N^2 - 1 ) }{k+ N} \,.
\ee
For large $N$ and fixed $k$ this scales linearly in $N$. It is worth
viewing this in terms of the 't Hooft coupling in the ABJM
model. It was demonstrated in \cite{Aharony:2008ug} that the effective 't Hooft coupling
of the theory is  $\lambda = {N \over k}$.

There are two natural limits to examine, large and small 't Hooft
coupling. In the large 't Hooft coupling limit,
\be \lambda>>1 \, \qquad c \rightarrow kN=k^2\lambda \ee
and in the small 't Hooft coupling limit where
\be \lambda<<1 \, \qquad c \rightarrow N^2=k^2\lambda^2 . \ee

As explained in the introduction, open membranes can end on five-branes.  The M5 world volume description of this is the self dual string soliton solution to the non-linear five-brane equations of motion.
One can calculate the absorbtion cross section of scalar fluctuations
of the five-brane equations of motion in the SDS background. This
indicates that the number of degrees of freedom of the SDS scales
linearly with the SDS charge
\cite{Berman:2001fs}
  or, equivalently, linearly with the number of M2 branes.  More subtle
anomaly considerations confirm this result \cite{Berman:2004ew}
and can also indicate how the degree of freedom count depends on the number of five-branes (for a review of all of this see \cite{Berman:2007bv}).

The linear scaling with $N$ of the central charge derived above
suggests we are on the right track in trying to interpret the boundary
theory of open membranes as describing $N$ coincident self dual
strings in the large 't Hooft coupling limit.

It remains a (realistic) calculational challenge to reproduce the $N^2$ scaling of
the boundary in the weak 't Hooft coupling theory.

Of course, the derivation of the WZW model is innately tied to the topological nature of the Chern-Simons theory.  Considering the full ABJM model makes such an interpretation much harder due to the decidedly non-topological matter sector.  Nevertheless, it seems that degrees of freedom associated to the self dual string may arise as the remnants of the would-be non-propagating pure gauge degrees of freedom in the membrane.

\section{${\cal N} = 1$ Supersymmetry with Boundary - General Theory}

We now move on to discuss the supersymmetric theory.
We will use supersymmetry to motivate the inclusion of certain
boundary terms  for the case of open membranes.
In particular, we shall try to build an action
that is automatically supersymmetric in the presence of a boundary.
This construction requires no boundary condition and thus holds
off-shell  i.e. without the imposition of Euler-Lagrange boundary
conditions.
We shall ultimately use the ${\cal N} = 2$ superspace form of the ABJM model introduced in \cite{Benna:2008zy}, but, before exploring that, we review the approach for some simpler models.  We first introduce a formalism employed by Belyaev and van Nieuwenhuizen, \cite{Belyaev:2008xk}, for ${\cal N}=1$ supersymmetry and apply this Chern-Simons matter theories.  We then develop this idea to tackle ${\cal N} = 2$ supersymmetry and the ABJM theory.

In \cite{Belyaev:2008xk} it is shown how to build 3-dimensional supersymmetric actions when space-time has a boundary and we now review this procedure.
A ${\cal N}=1$  scalar superfield is given by\footnote{See appendix for details of supersymmetry conventions.}
\bea
\Phi = a + \theta \psi - \theta^2 f \, ,
\eea
and can be integrated over superspace to form an action
\bea
S_0= \int d^3 x \int d^2 \th  \, \Phi =  \int d^3 x \, f\, .
\eea
The supersymmetry transformations,
\bea
\d \Phi = \e Q \Phi \Rightarrow  \left\{  \begin{array}{l} \d a = \e \psi\\
\d \psi_\a = -\e_\a f + (\c^\mu\e)_\a \p_\mu a\\
\d f = - \e\c^\mu \p_\mu \psi \, \, , \end{array} \right.
\eea
ensure that the action varies to a total derivative under rigid supersymmetry. When space-time has no boundary such terms can be safely ignored using generic arguments and the action is thus supersymmetric.
 In the case that space-time has a boundary, which we will assume
 throughout to be spatial and lie at $x_3 = 0$, we must pay attention
 to this surface term.  Without any other considerations we have
 broken supersymmetry as $\delta S_0 = -\partial_{\mu}(\epsilon \gamma^{\mu} \psi )$.

   One might stop here and say that supersymmetry is recovered by imposing some boundary conditions. However, we should be clear in distinguishing Euler-Lagrange boundary conditions which are associated with equations of motion  and the kind of boundary condition that are required to enforce supersymmetry off-shell.  Instead the approach advocated by \cite{Belyaev:2008xk} is to build actions that are supersymmetric without the need for any boundary conditions.  Only then, having built such a bulk + boundary supersymmetric action, should one go ahead and calculate the EL field equations and boundary conditions if so desired.

 Consider the following boundary action
 \bea
 S_1 = -\int d^3 x \p_3 \Phi |_{\th=0}  = -\int d^3x\p_3 a\,,
 \eea
with supersymmetry variation
 \bea
 \d S_1 = -\int d^3 x \,   \p_3 (\e \psi)\,.
 \eea
 Then the combination $S_0 \pm S_1$ has variation
 \bea
 \d [S_0 \pm S_1] = \mp \int d^3 x  \p_3[ \e (1 \pm \c^3)\psi] = \mp \int d^3 x  \p_3[ 2\e_\mp\psi_\pm]
 \eea
where we have defined projected spinors $\psi_\pm \equiv P_\pm \psi \equiv  \frac{1}{2}  (1 \pm \gamma^3)\psi$.
Then
\bea
\d [S_0 \pm S_1] = 0 \Leftrightarrow \e_\mp = 0  \,.
\eea
 Hence the modified action preserves half (${\cal N} = (1,0)$ or $(0,1)$) of the supersymmetry generated by $\e_\mp Q_\pm$.

We may augment this minimal process by including an extra ${\cal N} = (1,0)$ 2-dimensional theory defined solely on the boundary.  To this end it is helpful to relate 3-dimensional  ${\cal N} = 1$ multiplets to 2-dimensional ${\cal N} = (1,0)$ multiplets.  This is detailed in the appendix.

\section{${\cal N} = 1$ Super Chern-Simons with Boundary}

Let us apply this formalism to  ${\cal N} = 1$  abelian Chern-Simons theory. This is addressed in \cite{Belyaev:2008xk} and we recapitulate this here for convenience and to clarify a few subtleties which will become important in the more involved scenarios we consider later on.

We begin with a spinor superfield,
\bea
\Gamma_\a = \chi_\a - \th_a M  + (\c^\mu \th)_\a v_\mu - \th^2 [ \lambda + \c^\mu\p_\mu \chi]_\a \, ,
\eea
which contains the 3-dimensional vector field as one of its components.
The notion of gauge transformation is extended to superspace
\bea
\d_{G} \Gamma_\a = D_\a \Phi  \Rightarrow \left\{ \begin{array}{l}
\d_{G} \chi = \psi\\
\d_{G} M = f\\
\d_{G} v_\mu = \p_\mu a\\
\d_{G} \lambda = 0\, .
 \end{array}\right.
 \eea
 When an action is invariant under this gauge transformation the arbitrary shifts in $M$ and $\chi$ allow the WZ gauge choice  $M = \chi = 0$.
 The gauge invariant field strength is given as
 \bea
 W_\b = D^\a D_\b \Gamma_\a = \lambda_\b  + 2 \e^{\mu\nu\rho} ( \th \c_\rho)_\b \p_\mu v_\nu + \th^2(\c^\mu \p_\mu  \lambda)_\a \,.
 \eea
 The Chern-Simons action is given by
 \bea
 \label{SCS0}
 S^{CS}_0 &=& \int d^3 x \int d^2 \th \, \Gamma^\a W_\a \\
 &=& \int d^3 x \, \lambda \lambda  - 4\e^{\mu\nu\rho}  v_\mu \p_\nu v_\rho  - \p_\mu (\chi \gamma^\mu \lambda)\,.
 \eea
Notice that the auxiliary field $M$ is entirely absent and that $\chi$ enters only as a total derivative.   This action is gauge invariant only up to a total derivative
\bea
\d_{G} S^{CS}_0 = \int d^3 x \p_\mu[ \lambda \c^\mu \psi + 4 \e^{\mu \nu \lambda} \p_\nu a  v_\lambda]\,.
\eea
When we have a boundary we must be careful about such terms.  In keeping with our overall philosophy we do not impose a boundary condition just to restore a symmetry.  For the moment we take the view point that we have destroyed the gauge symmetries
\bea
\d_{G} \chi = \psi \, , &  \d_{G} v_m = \p_m  a \, ,
\eea
where $m=\{0,1\}$.  The symmetry associated to the absent field $M$ remains (trivially) as does that corresponding to $v_3$.  The supersymmetry is also destroyed.

We now follow the procedure of the preceding section and add a supersymmetric restorative term
\bea
\label{SCS1}
S^{CS}_1= - \int d^3 x  \p_3 (\Gamma^\a W_\a )|_{\theta=0}= -\int d^3 x  \p_3 (\chi^\a \lambda_\a )\,
\eea
Then
\bea
S^{CS}_0+ S^{CS}_1 =  \int d^3 x \, \lambda \lambda - 4\e^{\mu\nu\rho}  v_\mu \p_\nu v_\rho     - \p_3 ( 2 \lambda_+  \chi_-)
\eea
is, by construction, invariant under $\e_+ Q_-$ supersymmetry transformations.  Notice that by the elimination of $\chi_+$ we have also restored some gauge symmetry namely
\bea
\d_{G} \chi_+ = \psi_+ \,.
\eea
Unfortunately this modified action is a little awkward since the non-propagating gaugino has a coupling on the boundary.  This prevents us from straightforwardly integrating out the gaugino which is something we may wish to do when we couple to matter.  We would like to remove this term  without resorting to an ad-hoc off-shell boundary condition on $\lambda$ and thereby voiding the construction of supersymmetry without boundary conditions.

We are at liberty to supplement this construction with any two-dimension ${\cal N} = (1,0)$  theory defined on the boundary.  Suppose this system is built from the same fields we already have.
Then we can find appropriate actions by decomposing $\Gamma_\alpha$ into co-dimension 1 superfields as is detailed in the appendix.
 One finds that
\bea
\hat{\Gamma}^-_{\a} &=&  {\chi}_{- \a} + (\c^m\th_+)_\a v_m \\
\hat{\Sigma}^+_m &=&   v_m  + \th_{+} [\frac{1}{2}\c_m \lambda_+ +
\p_m \chi_-]  \label{boundarydef}
\eea
are two such 2D N=(1,0) superfields. From these we can form the boundary action
\bea
\label{SCS2}
S^{CS}_2 &=& -2\int d^3 x  \p_3 \big\{ \int d\th^\a_{+} \c^{m\b}_\a \hat{\Gamma}^{-}_\b \hat{\Sigma}^+_m  \big\} \\
&=& 2\int d^3x \p_3 [ \chi_- \lambda_+ + \chi_- \c^m \p_m \chi_- + v_m v^m ]\,.
\eea
 We combine this with our previous expressions to find a total action for Chern-Simons with a boundary
\bea
\label{N=1CStot}
S^{CS}_{tot} = S^{CS}_0+ S^{CS}_1 + S^{CS}_2  = \int d^3x \,  \lambda \lambda  -4\e^{\mu\nu\rho}  v_\mu \p_\nu v_\rho    + 2\p_3 [ \chi_- \c^m \p_m \chi_- + v_m v^m ]\,
\eea
which preserves $\e_+$ supersymmetry and which contains a non-propagating gaugino which can be easily integrated out. It is interesting to note the appearance of a dynamical fermion on the boundary. This field, which would have been pure gauge in the WZ sense, has been promoted to become dynamical.  This sort of behavior is analagous to the purely bosonic case we looked at before. It should be noted that the kinetic term for this field is really chiral; one simply makes use of the identity $\c^0\c^1=\c^3$ to show
\be
\chi_- \c^m \p_m \chi_-  = \chi_- \c^1 \p_+ \chi_- \,.
\ee
The new combined action has gauge transformation
\bea
\label{Gaugevariation}
\d_{G} S^{CS}_{tot}  &=& \int d^3 x   4 \p_3[ \psi_- \gamma^m \p_m \chi_- + (\eta^{mn} + \e^{mn} ) \p_m av_n ] \\
&=& \int d^3 x   4 \p_3[ \psi_- \gamma^m \p_m \chi_- -  \p_+ a v_-]\, ,
\eea
where we have defined light cone combinations $v_\pm = v_0 \pm v_1$.
So for the final action the full set of gauge symmetries are
\bea
\delta  v_-  = \p_- a\, , &\,& \delta  v_3  = \p_3 a\,,\\
\delta M = f \, , &\,&    \delta \chi_+ = \psi_+ \, ,
\eea
(the last two are somewhat trivial since those fields are absent).

In summary we started with $S_0$, (\ref{SCS0}), an action that was neither gauge invariant nor supersymmetric in the presence of a boundary.  We added a suitable term $S_1$, (\ref{SCS1}) to restore $\e_+$ susy.  We then added a separate boundary action $S_2$, (\ref{SCS2}) which in itself is $\e_+$ supersymmetric.  The final result is a combined action, $S^{CS}_{tot} = S_0 + S_1 + S_2$, (\ref{N=1CStot}), which preserves half the supersymmetry and \textquoteleft half\textquoteright\, the gauge symmetry.
Since $\chi_-$ appears in the final theory as a propagating boundary field it seems that we might have to be cautious about adopting Wess-Zumino gauge.  We shall return to this point when we discuss couplings to matter.

\subsection{EL boundary conditions and WZW model revisited}

We are now in a position to perform an Euler Lagrange variation of the total action.  We find the bulk variation yields the usual equations of motion
\bea
\lambda = \e^{\mu\nu\rho}\p_\mu v_\rho=0
\eea
and the boundary variation requires
\bea
\left( v_+\delta v_- + \delta \chi_- \c^m \p_m \chi_- \right)_{\p M} = 0\,.
\eea
With respect to the transverse  direction we view this equation as
having the form $p\delta q$.  Neumann conditions are of the form $p=0$
and Dirichlet are $q=const$. The correct supersymmetric boundary
conditions can be easily read off from the boundary ${\cal N}= (1,0)$
multiplets we have used earlier (\ref{boundarydef}).  The Neumann
condition is, given in terms of those multiplets, $\hat{\Sigma}^+_{m=+}=0$ and the Dirichlet is $\hat{\Gamma}^-_\a= const$. For the gauge field these are just $v_+=0$ for Neumann and $v_- = 0$ for Dirichlet\footnote{The Dirichlet condition on a superfield allows the lowest component to be a non-zero constant but other components must be zero.}.

Even in the bosonic sector the $v_mv^m$ boundary term  in (\ref{N=1CStot}) is a departure from the standard CS theory.  This has an implication for the boundary model obtained.  If we compare to the discussion in section 2, we see that we have a different boundary condition.  Instead of $v_0 = 0$, we may choose $v_+ = v_0 + v_1 =0$.  In the bosonic action
\bea
S_{bos} = \int_M d^3x \, -4\e^{\mu\nu\rho}  v_\mu \p_\nu v_\rho    + 2\p_3 [ v_m v^m ]\
\eea
we may make use of the boundary condition and carry out some
integration by parts to obtain
\bea
S_{bos}  = -4\int_Md^3x \, \e^{0ij} \left(v_0 (\p_i v_j - \p_j v_i)  - v_i \dot{v_j} + \p_j(v_iv_0) \right)
\eea
with $x^i=\{x^1, x^3\}$.  In the standard derivation of the boundary
WZW action one would now use the b.c. to eliminate the total
derivative term in the above, and having done so,
$v_0$ becomes a simple Lagrange multiplier.
Although we can not do exactly this we can still use the boundary
condition to modify the total derivative term by replacing $v_0$ with $-v_1$ on the boundary.  The action is equivalent to
\bea
S_{bos}  = -4\int_M d^3x \,\e^{0ij} \left(2v_0 \p_i v_j  - v_i \dot{v_j} - \p_j(v_iv_1) \right)\,.
\eea
We can now use the field equation for $v_0$ to invoke the pure gauge constraint $v_i = \p_i U$.  Plugging into the action we are left with
\bea
S_{bos}  &=& -4\int_M d^3x \e^{0ij} \left( -\p_j( \p_i U \p_0 U) - \p_j(\p_iU \p_1 U)  \right)\\
&=& 4\int_{\p M} dx^0 dx^1 \left( \p_1 U \p_0 U + \p_1 U \p_1 U)  \right)\,.
\eea
The result appears surprising since manifest Lorentz symmetry appears to be lost and the kinetic term looks somewhat unconventional.   However this sort of  two-dimensional action is not unknown; it is the Floreanini Jackiw (FJ) action \cite{Floreanini:1987as} for a chiral field.  The field equations show that $\p_0\p_1 U = -\p_1 \p_1 U$, and so after integrating\footnote{The arbitrary function that arises upon integration is set to zero by invoking suitable boundary conditions.} one sees that $U$ is indeed a chiral boson. The boundary Majorana-Weyl fermion term is the natural superpartner for this chiral boson.

 Although we have only addressed the abelian CS theory it seems very plausible that similar considerations in the non-abelian context would result in the supersymmetric chiral WZW of Sonnenschein \cite{Sonnenschein:1988ug};
    the kinetic term becomes FJ like, the Wess-Zumino term is unaltered and the super-partner is a free adjoint Majorana-Weyl fermion.

 We now comment on how this might generalize to the ABJM model.
 First let us think about the bosonic sector.  We saw in section 2
 that we should anticipate a non-chiral WZW model for the ABJM theory.
 However we have changed the boundary condition used in the derivation
 of the WZW model.    To construct a non-chiral boundary string theory
 we must choose boundary conditions that do not break the parity
 invariance.  Therefore the appropriate parity preserving boundary conditions are $v_+ =0$ for one gauge field and $\hat{v}_- = 0$  for the other.  In the bosonic sector this choice would indeed result in chiral and anti chiral FJ action for a boundary boson.

  However, such a choice is not compatible with preserving ${\cal N }= (1,0)$ supersymmetry.  Our bulk + boundary construction applied to the case of two gauge group factors would yield two propagating boundary fermions with the same chirality. This is quite clearly not the correct supersymmetric completion of two chiral bosons of opposite chirality. Of course, this is really a triviality. Breaking half of ${\cal N}= 1$ supersymmetry in necessarily chiral.

\section{${ \cal N}= 1$ Super Chern-Simons Matter Theory}
We begin with a pure matter theory based on a real scalar superfield $\Phi=(\phi,\psi,f)$.
The action is given as
\bea
S^M_0 &=&\int d^3 x \int d^2 \th   -\frac{1}{2} D^\a \Phi D_\a \Phi\\
&=&  \int d^3 x ff - \p_\mu \phi \p^\mu \phi -  \psi \c^\mu \p_\mu \psi
\eea
and the extra contribution
\bea
S^M_1 = \int d^3 x \p_3 [\frac{1}{2} D^\a \Phi D_\a \Phi]_{\th=0} =  \int d^3 x \p_3 \left(\psi_+ \psi_- \right)
\eea
is such that $S_0+S_1$ has $\e_+$ SUSY.  Since no auxiliary field occurs on the boundary there is no particular motivation to add any additional boundary action to this. We now consider the $U(1)$ gauged version of this action for complex fields with a Chern-Simons kinetic term for the gauge field. The two matter terms to consider are
\bea
S^M_0 &=& \int d^3 x \int d^2 \th   -\frac{1}{2} (D^\a+ i\Gamma^\a) \Phi^\ast ( D_\a - i\Gamma_\a) \Phi \,,\\
S^M_1 &=& \int d^3 x \p_3 \big[ \frac{1}{2} (D^\a+ i\Gamma^\a) \Phi^\ast ( D_\a - i\Gamma_\a) \Phi \big]_{\th=0}\,.
\eea
Both of these terms are completely invariant under the superspace generalization of gauge symmetry:
\bea
\Gamma_\a &\rightarrow& \Gamma_\a + D_\a\Lambda\, ,\\
\Phi &\rightarrow& \exp (i\Lambda) \Phi\, .
\eea
By construction $S_0+S_1$ is now $\e_+$ supersymmetric. At this stage we must think a little about how we expand these actions into components.  One would normally adopt Wess-Zumino gauge ($\chi = M = 0$) however, we have established that $\chi_-$ occurs as a propagating boundary field in the Chern-Simons kinetic term and the WZ symmetry associated to this field is broken.  We ask to what extent can we work in WZ gauge when evaluating the above actions?

Let us answer this question in the abstract.  Suppose we have some fields donated by ${\chi, \phi}$ and some symmetry $U$ which acts as a shift on $\chi$ i.e.
\bea
U: \left\{ \begin{array}{l}
\chi \mapsto \chi^U = \chi + U\\
\phi \mapsto \phi^U
 \end{array}\right.
\eea
For an action $S[\chi,\phi]$ that is invariant under these
transformations we can gauge fix in the standard way. We introduce a
fiducial choice $\hat{\chi} = 0$ and insert this into the path integral;
the Fadeev Popov determinant is trivial, and gauge invariance allows
us to perform the integration over the gauge group\footnote{One must also check if there are any constraints that we must enforce on the Hilbert space due to  \textquoteleft missing\textquoteright\, equations of motion (for example, in string theory fixing conformal gauge requires that the stress tensor must act as zero on the Hilbert space).  In the cases that we are interested in however the remaining equations of motion after gauge fixing automatically imply the \textquoteleft missing\textquoteright\, equation from WZ fields.}.

 Now let us suppose that there is an addition action $S'[\chi]$ that
 is not invariant under the symmetry $U$.  This is exactly the
 situation we have found ourselves in. Let us go on regardless with
 the gauge fixing procedure and see where we end up. This time we are
 not able to perform the integration over the gauge group. It may
 seems as though we have not achieved anything but, the key point is
 that for the portion of the action $S$ that does not break the
 symmetry we can adopt the gauge fixing choice.  Applied to our
 Chern-Simons matter theory this shows that weare allowed to consider the matter sector in the WZ gauge\footnote{Of course, there are the regular issues of the WZ gauge breaking supersymmetry but this is not pertinent to our discussion.}.

Expanding in components we find that the total $N=(1,0)$ Chern-Simons matter theory with a boundary is given by\footnote{We have included a normalisation factor $\kappa$ for the Chern-Simons term.  In the non-abelian theory, invariance under large gauge transformations requires $\kappa$ to obey a quantisation condition.  With traces normalised so that $Tr(T^a T^b) =\delta_{ab}$ we have that $\kappa \in \frac{\mathbb{Z}} {32\pi}$\,. }
\bea
S^{CS-M}_{tot}&=& S^{CS}_{tot} + S^{M}_0 + S^{M}_1\\
 &=& \int d^3x \,  \kappa \lambda \lambda  - 4\kappa\e^{\mu\nu\rho}  v_\mu \p_\nu v_\rho     +  ff^\ast - \nabla^\mu \phi \nabla_\mu \phi^\ast - \psi^\ast \c^\mu \p_\mu \psi \\
&&\quad + \frac{i}{2} (\psi \lambda \phi^\ast  -  \psi^\ast \lambda  \phi)\\
&&\quad +\p_3 [ 2\kappa\chi_- \c^m \p_m \chi_- +2\kappa v_m v^m  + \psi^\ast_- \psi_+]\, .
\eea
Now the importance of removing the boundary gaugino couplings becomes clear; we can integrate out the gaugino to generate the potential.  The result (generalized to include capital Roman flavor indices) is given by
\bea
S^{CS-M}_{tot}&=& \int d^3x \,   - 4\kappa\e^{\mu\nu\rho}  v_\mu \p_\nu v_\rho  \e^{\mu \nu \lambda}  - \nabla^\mu \phi_I \nabla_\mu \phi_I^\ast - \psi^\ast_I \c^\mu \p_\mu \psi_I\\
&&\quad + \frac{1}{16\kappa}\left(\psi^\ast_I \psi^\ast_J \phi_I\phi_J +\psi_I \psi_J \phi^\ast_I\phi^\ast_J  -2 \psi^\ast_I \psi_J \phi_I\phi^\ast_J \right)\\
&&\quad +\p_3 [ 2\kappa\chi_- \c^m \p_m \chi_- +2\kappa v_m v^m  + \psi^\ast_{-I} \psi_{+I}]\, .
\eea
In this case the boundary terms are not especially interesting;
they are just the combination of the ones obtained for Chern-Simons
theory and ungauged matter separately.

\section{${ \cal N}= 2$ Supersymmetry with Boundary - General Theory}

We now move up in complexity by considering extended supersymmetry.
In three-dimensions ${\cal N}=2$ superspace is realized by taking the Grassman coordinates to be complex:
\begin{equation}
\th_\a = \frac{1}{\sqrt{2}} \left(\th_{1\a} + i \th_{2\a}\right)
\end{equation}
  We wish to generalize the procedure used in the ${\cal N}=1$ case to restore supersymmetry with boundaries.  It is easiest to do this by working in a basis  where we decompose the ${\cal N}=2$ supersymmetry into two copies of the real ${\cal N}=1$ symmetry \cite{Ivanov:1991fn} with the algebra\footnote{See appendix for details.}
\bea
\{Q_i\, , Q_j\} = 2\delta_{ij}\gamma^\mu \p_\mu & i=\{1,2 \} \,.
\eea
Under this decomposition the supersymmetry parameters are two 2-component Majorana spinors $\epsilon_1^\alpha$ and $\epsilon_2^\alpha$.  Thus we may repeat the construction for ${\cal N}=1$ SUSY with a boundary twice.  We start with the most general  ${\cal N}=2$  action
\bea
S_0 =- \int d^3 x d^2 \th d^2 \bar{\th} \, L[\th, \bar{\th}] =  \int d^3 x d^2 \th_1 d^2 \th_2 \, L[\th_1, \th_2]\, .
\eea
We augment this action with another term
\bea
S_1 = -\int d^3 x d^2 \th_1  \p_3 L|_{\th_2=0}\,,
\eea
so that $S_0 + S_1$ has the $\e_{2+}Q_{2-}$ supersymmetry.
There is now an apparent choice for the remaining supercharge as to whether we preserve the same chirality supersymmetry $\e_{1+}Q_{1-}$ or the opposite  $\e_{1-}Q_{1+}$. We consider two further terms
\bea
S_2 &=& -\int d^3 x \int d^2 \th_2\, \p_3 L|_{\th_1 =0}\, ,\\
S_3&=& \int d^3 x \,  \p_3\p_3L|_{\th_1 = \th_2 = 0}
\eea
Then $S_0+S_1+S_2+S_3$ preserves $(\e_{1+}, \e_{2+})$ supersymmetry and we shall describe this choice as  ${\cal N}=(2,0)$.  This can be expressed as a projection condition on the complex spinor
  \begin{equation}
  P_+ \epsilon = P_+ \frac{1}{\sqrt{2}}\left(\e_1 + i \e_2  \right) = \epsilon\,.
  \end{equation}

   On the other hand $S_0+S_1-S_2-S_3$ preserves $(\e_{1-}, \e_{2+})$
   supersymmetry and we shall
call this choice  ${\cal N}=(1,1)$.

In  ${\cal N}=2$ theories we can also have superpotentials of the form
\bea
S^W = \int d^3x\int d^2 \th W(\Phi) +  c.c.
\eea
where $\Phi$ is an ${\cal N}=2$ chiral superfield.  We apply the above procedure and find
\bea
S^W_0&=&\int d^3x d^2\th W(\Phi) =  \int d^3x d^2 \th d^2 \bar{\th} (- \bar{\th}^2)   W(\Phi)\\
&=& \int d^3 x d^2 \th_1 d^2 \th_2 \frac{1}{2}(\th_1^2-\th_2^2 - i \th_1 \th_2)W(\Phi)\,
\eea
and
\bea
S^W_1 &=& \int d^3 x d^2 \th_1 \, \p_3 [- \frac{1}{2}(\th_1^2-\th_2^2 - i \th_1 \th_2)W(\Phi) ]_{\th_2=0} = \int d^3 x \, \frac{1}{2}\p_3 W(a)\, ,\\
S^W_2&=& \int d^3 x d^2 \th_2 \, \p_3 [- \frac{1}{2}(\th_1^2-\th_2^2 - i \th_1 \th_2)W(\Phi) ]_{\th_1=0} = -\int d^3 x \, \frac{1}{2}\p_3 W(a)\, ,\\
S^W_3 &=& -\int d^3 x \,   \p_3\p_3 [- \frac{1}{2}(\th_1^2-\th_2^2 - i \th_1 \th_2)W(\Phi)]_{\th_1 = \th_2 = 0}=0\,.
\eea
In which $a$ is the lowest component of $\Phi$.
The implication is that if we are to preserve ${\cal N}=(2,0)$ supersymmetry a superpotential term requires no boundary contribution however to preserve ${\cal N}=(1,1)$ supersymmetry we must add to the lagrangian a boundary term of $\p_3 W(a)$.

\section{${ \cal N}= 2$ Chern-Simons with Boundary}
The abelian ${ \cal N}= 2$ Chern-Simons  theory is described by an action \cite{Ivanov:1991fn,Avdeev:1991za}
\bea
S_0= \int d^3 x d^2\th d^2 \bar{\th} VD^\a \bar{D}_\a V\,.
\eea
The vector superfield $V$ can be expanded in to ${ \cal N}= 1$ component superfields as
\bea
V(\th_1, \th_2) = A(\th_1) + \th_2 \Gamma (\th_1) - \th_2^2 (B(\th_1) - D_1^2A)
\eea
with components summarized by
\bea
A=(a, \psi, f) \, ,& B=(b, \eta, g) \, , & \Gamma = (\chi, M, v, \lambda)\, .
\eea
The extended supersymmetric gauge transformation allow the \textquoteleft Ivanov\textquoteright\, gauge  choice whereby we set $A=0$ and invoke standard WZ gauge for the spinor multiplet namely $\Gamma = (0, 0, v, \lambda)$ \cite{Ivanov:1991fn}.  Note that this choice differs from the gauge choice adopted in some of the rest of the literature e.g. \cite{Gaiotto:2007qi, Benna:2008zy}.

We expand $S_0$  into ${ \cal N}= 1$ superfields
\bea
S^{CS}_0 =  \int d^3 x d^2\th_1 (BB+\Gamma^\a W_\a + \frac{1}{2}D_1^\a(D_{1\a} B  A  -  B D_{1\a}  A))\, .
\eea
In this action the components of A occur only inside a total derivatives which can easily been seen from the identity $D^2D^\a = (\c^\mu D)^\a \p_\mu$.
Thus, without a boundary, the only  difference between this and ${\cal N}=1$ Chern-Simons theory is the appearance of an auxiliary multiplet. In components
\bea
S^{CS}_0&=& \int d^3 x 2gb +\eta \eta + \lambda \lambda - 4\e^{\mu \nu \rho}v_\mu \p_\nu v_\rho \\
&& \quad \quad +\p_3 (  \lambda\c^3\chi  +\eta\c^3 \psi   +\p^3b a - b\p^3a)\, .
\eea
Following the rules of section 6, we build the extra terms
\bea
S^{CS}_1&=& \int d^3 x \p_3(-ag - bf - \eta \psi)\, ,\\
S^{CS}_2&=& \int d^3 x \p_3( - bb - \lambda \chi - ag + bf) \, ,\\
S^{CS}_3&=& \int d^3 x \p_3( \p_3 a b + a \p_3 b)\, .
\eea

Hence we form a ${\cal N}=(2,0)$ action
\bea
S^{CS}_{(2,0)} &=& S^{CS}_0+S^{CS}_1+S^{CS}_2+S^{CS}_3 \\
 &=& \int d^3 x 2gb +\eta \eta + \lambda \lambda - 4\e^{\mu \nu \rho}v_\mu \p_\nu v_\rho \\
&& \quad \quad + \p_3 ( -2\chi_- \lambda_+ - 2\psi_-\eta_+ - bb - 2a(g-\p_3b) )\, ,
\eea
and a ${\cal N}=(1,1)$ action
\bea
S^{CS}_{(1,1)} &=& S^{CS}_0+S^{CS}_1-S^{CS}_2-S^{CS}_3 \\
 &=& \int d^3 x 2gb +\eta \eta + \lambda \lambda - 4\e^{\mu \nu \rho}v_\mu \p_\nu v_\rho \\
&& \quad \quad + \p_3 ( 2\chi_+ \lambda_- - 2\psi_-\eta_+ + bb - 2(f+\p_3a)b) \, .
\eea
In both of these cases we find that we have a gaugino coupling on the boundary.  Motivated by the ${\cal N}=1$ example we now construct some 2d boundary terms that can be used to remove this term.

For the ${\cal N}=(2,0)$ case we consider the addition of
\bea
S^b_{(2,0)} = \int d^2x \int d\th_{1+}\c^m d\th_{2+} 2\hat{\hat{V}}\hat{\hat{V}}_m \, ,
\eea
where the half supersymmetric superfields are defined and constructed in the appendix. Expanding out into components one finds that
 \bea
 S^{CS}_{(2,0)}-S^b_{(2,0)} &=&  \int d^3 x  2gb +\eta \eta + \lambda \lambda - 4\e^{\mu \nu \rho}v_\mu \p_\nu v_\rho \\
&&  \quad + 2\p_3 ( v_nv^n + \chi_- \c^m\p_m \chi_- +  \psi_- \c^m\p_m \psi_-+ \p_ma\p^m a  - \frac{1}{2} bb)\,.
 \eea
 In this case we have removed both the gaugino couplings from the boundary at the expense of introducing some propagating boundary fields.  We have two propagating fermions of the same chirality; this indicates the non-chiral $(2,0)$  nature of the symmetry.

 Also notice that we have not eliminated the auxiliary field $b$ from the boundary. This will actually prove to be a source of interest to us.  Notice that the other auxiliary scalar $g$ serves as a Lagrange multiplier enforcing $b$ to take a particular value.  We then have no need to eliminate $b$ by its equations of motion and so it is not a problem that it enters in our boundary term.   Moreover, when we couple to matter this will provide  an interesting boundary interaction term for the matter fields.

For the ${\cal N}=(1,1)$ case we consider
\bea
S^b_{(1,1)} = \int d^2x \int d\th_{1-} d\th_{2+}2 \tilde{ \hat {U}}^\a \tilde{\hat{V}}_\a
\eea
where again we refer the reader to the appendix.  We find that
\bea
S^b_{(1,1)} = -\int d^3x 2\p_3[\chi_+ \lambda_-  +  \psi_-\eta_+  +  \chi_+ \c^m\p_m \chi_+ +  \psi_- \c^m \p_m \psi_- + b(f+\p_3a) - (f-\p_3a)^2]\,.
\eea
Unlike the ${\cal N} = (2,0)$ case we can no longer eliminate both the gaugino boundary couplings.  The best we can do is
 \bea
 S^{CS}_{(1,1)}-S^b_{(1,1)} &=& \int d^3 x 2gb +\eta \eta + \lambda
 \lambda - 4\e^{\mu \nu \rho}v_\mu \p_\nu v_\rho  \\
&&  + 2\p_3 ( v_nv^n +  \chi_+ \c^m\p_m \chi_+ + 2 \chi_+\lambda_- +
\psi_- \c^m\p_m \psi_- - (f+\p_3a)^2 + \frac{1}{2} bb)\,. \nonumber
 \eea
 On the boundary the two fermions of opposite chirality seems to indicate the $(1,1)$ nature of the theory.

 We see that the  $\chi_+\lambda_-$ interaction remains. This is not a necessarily a problem; it simply means that the auxiliary field is needed for off-shell supersymmetry without boundary conditions.  If we are interested in on-shell effects we may choose a boundary condition to eliminate this term providing it is compatible with the preserved supersymmetry. Such a boundary condition is be encoded in the (1,1) multiplets.  For instance we may choose as a boundary condition for the gauge sector
 \bea
 \label{boundarycondition}
 0= \tilde{\hat{U}}_\a &=& \chi_{+\a} - (\th_{1-}\c^m)_\a v_m   - \th_{2+\a} (b-(f+\p_3a))+ \th_{2+}\th_{1-}(\eta_+ +\c^m\p_m \psi_-)
\eea
This choice of boundary condition suggests that $\chi_+ = 0$.      This boundary condition would certainly eliminate any concerns about the $\chi_+\lambda_-$ interaction; on-shell we could freely integrate out $\lambda$ from the bulk.  It also suggests that $\chi_+$ may not actually be a propagating degree of freedom at all.

We also have a non-propagating scalar squared term on the boundary given by $(f+\p_3a)^2$.  A naive Euler Lagrange variation would suggest setting this term to zero, at least on-shell.  However we can see form the boundary multiplet (\ref{boundarycondition}) that compatibility with supersymmetry also requires that $b=0$.  Similarly we see that $\psi_-$ obeys a simple 2d fermion equation of motion provided that $\eta_+ = 0$ on the boundary.

\section{${ \cal N} = 2$ Matter with Boundary}

${ \cal N} = 2$ matter is describe by chiral and anti chiral superfields with ${ \cal N} = 1$ expansion
\renewcommand{\Z}{{\cal Z}}
\newcommand{\W}{{\cal W}}
\bea
\Z(\th_1, \th_2)  &=& \frac{1}{2}Z(\th_1) + \frac{1}{2}i\th_2D_{1\a} Z + \frac{1}{2}\th_2^2 D_1^2 Z \, ,\\
\bar{\Z}(\th_1, \th_2) &=& \frac{1}{2}Z^\ast (\th_1) - \frac{1}{2}i\th_2D_{1\a} Z^\ast + \frac{1}{2}\th_2^2 D_1^2 Z^\ast\, ,
\eea
where the factors of half are for later convenience and the components of the ${ \cal N} = 1$ superfields are summarized by $Z(\th_1) = (Z, \xi, F)$.
With our conventions the kinetic term for gauged matter is given by
\bea
S_0 &=& \int d^3x d^4 \th \bar{\Z}e^{2V} \Z\, .
\eea
According to our earlier consideration of gauge fixing we evaluate this action using \textquoteleft Ivanov\textquoteright\, gauge
\bea
V(\th_1, \th_2) = \th_2 \Gamma (\th_1) - \th_2^2 B(\th_1)\, ,
\eea
with $\Gamma = (0,0,v_\mu, \lambda)$ and $F=(b, \eta, g)$.

First we consider the ${\cal N}=(2,0)$ case.  The construction of section 6, yields the following bulk + boundary action
\bea
S^{M}_{(2,0)} = \int d^3 x {\cal L}_{bulk} + \p_3 [ \xi_-^\ast \xi_+ + \frac{1}{2} Z^\ast b Z ]
\eea
where the bulk lagrangian is given by
\bea
  {\cal L}_{bulk} &=& F^\ast F -  \nabla_\mu Z^\ast  \nabla^\mu Z - \xi^\ast \c^\mu \nabla_\mu \xi \\
&& \, -\frac{1}{2}\left(Z^\ast b F + Z^\ast (\eta + i\lambda) \xi\right) + c.c.\\
 &&\, -\frac{1}{2} Z^\ast g Z -\frac{1}{2}\xi^\ast b \xi\,.
\eea
We observe that in this case we have a new gauge-matter coupling on the boundary of the form $ Z^\ast b Z $.  This will lead to new boundary interactions once the auxiliary field $b$  is eliminated.
If we now couple this matter sector to the ${\cal N}=(2,0)$  Chern-Simons term, generalize to include flavor indices and integrate out auxiliary fields, we find
\bea
S^{CSM}_{(2,0)} &=& \kappa S^{CS}_{(2,0)\, tot}+ S^{M}_{(2,0)}\\
 &=& \int {\cal L}_{kin} +  {\cal L}_{int} +{\cal L}_{pot} + \p_3  {\cal L}_{(2,0) \, bound}
 \eea
where
\bea
{\cal L}_{kin} &=& - 4\kappa \e^{\mu \nu \rho}v_\mu \p_\nu v_\rho - \nabla_\mu Z^\ast_A  \nabla^\mu Z^A - \xi_A^\ast \c^\mu \nabla_\mu \xi^A\\
{\cal L}_{int} &=& -\frac{1}{8\kappa}\left(2\xi^\ast_AZ^A Z^\ast_B \xi^B + \xi^\ast_A \xi^A Z^\ast_B Z^B \right)\\
{\cal L}_{pot} &=&  -\frac{1}{64\kappa^2}\left(Z^\ast_A Z^A \right)^3\\
{\cal L}_{(2,0) \, bound} &=&   2\kappa \left[ v_nv^n + \chi_- \c^m\p_m \chi_- +  \psi_- \c^m\p_m \psi_- +a\p_m\p^m a  \right] \\
&& \,\, +\xi^\ast_{A-} \xi^A_+  +\frac{1}{16\kappa}Z_A^\ast Z^A Z_B^\ast Z^B
\eea

The most striking new feature is the emergence of a scalar potential
on the boundary.  One possible interpretation for such a term can be
found in the classical literature \cite{Courant}.  Consider a classical membrane whose boundary is attached to the equilibrium displacement by means of zero natural length springs as displayed in figure 1.

\begin{figure}[!ht]
\label{Figure1}
\begin{center}
\scalebox{0.5}{\includegraphics{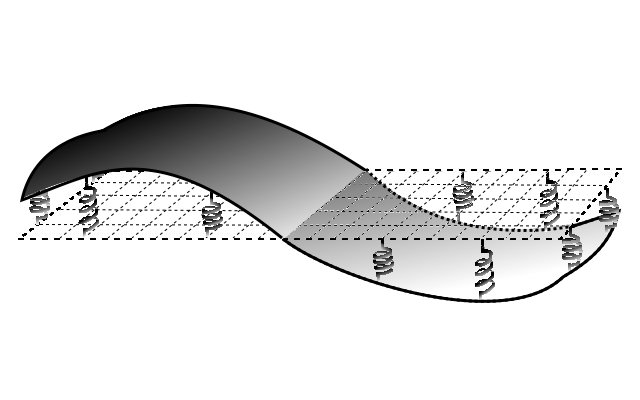}}
\end{center}
\caption{\em A membrane attached via springs at its boundary giving rise to boundary potential}
\end{figure}

  It is clear that such a system requires the inclusion of a boundary
  potential due to the potential energy stored in the elastic
  displacement of these springs.  Applying this thinking to our above
  action suggests that the boundary can be thought of as being
  attached to some sort of elastic material which displays a
  non-linear restorative force to displacement. We interpret this as
  being due to the fivebrane. Futher understanding of this term from
  the fivebrane perspective would be very desirable.

 If one now performs an Euler Lagrange variation we find a boundary condition for the scalar
 \bea
 \p_3 Z^A - \frac{1}{8\kappa} Z^A Z_B^\ast Z^B\, =0 .
 \eea
 How should we interpret this?

Let us consider searching for 1/2 supersymmetric bosonic vacuum solutions of the closed membrane theory and forget for a moment about the boundary terms.  This is most easily done by looking at the Hamiltonian and employing the Bogomolny trick.   We demand that the scalar fields are in a static configuration and only vary in the $x^3$ direction and the gauge fields are unexcited.  Then the Hamiltonian is given by
\bea
H &=& \p_3 Z^A \p_3 Z^\ast_A + \frac{1}{64\kappa^2}\left(Z^\ast_A Z^A \right)^3 \\
&=& |\p_3 Z^A - \frac{1}{8\kappa}Z_BZ^\ast_B Z^A|^2 + \frac{1}{16\kappa} \p_3(Z^\ast_A Z^A Z^\ast_B Z^B)\,.
\eea
 Then the minimum energy configuration satisfies the BPS bound
 \bea
 \p_3 Z^A - \frac{1}{8\kappa} Z^A Z_B^\ast Z^B\, =0 .
 \eea
 So we see that our \textquoteleft natural\textquoteright\, boundary condition obtained from the generalized theory corresponds exactly to the BPS equation.

  We now turn to the ${ \cal N} =(1,1)$ case. Here things are slightly different.  In the gauged matter sector we find
\bea
S^{M}_{(1,1)} = \int d^3 x {\cal L}_{bulk} + \p_3 [ \frac{1}{2} \xi^\ast\c^3 \xi + \frac{1}{2} F^\ast Z + \frac{1}{2}  Z^\ast F + \frac{1}{2} \p_3(Z^\ast Z) ]
\eea
where the bulk Lagrangian is unchanged.   We observe that the non-propagating scalar $F$ appears on the boundary.  Unlike the similar situations encountered before it seems to be impossible to eliminate all of these terms from the boundary by the addition of an extra boundary action.   This established by a detailed examination of the  ${ \cal N} =(1,1)$  boundary multiplets described in the appendix.   This means that the auxiliary field is required for the action to have supersymmetry without boundary conditions.

The full result for the ${ \cal N} =(1,1)$ Chern-Simons Matter theory is
\bea
S^{CSM}_{(1,1)} &=& \kappa S^{CS}_{(1,1)\, tot}+ S^{M}_{(1,1)}\\
 &=& \int {\cal L}_{kin} +  {\cal L}_{int} +{\cal L}_{F} + \p_3  {\cal L}_{(1,1) \, bound}
 \eea
where the kinetic and bose-fermi interaction terms are as before. The occurrence of $F$ on the boundary prevents simply integrating out $F$ and so we   have
\bea
{\cal L}_{F} = F^\ast F - \frac{1}{8\kappa}\left(F^\ast(Z^\ast Z)Z +  Z^\ast(Z^\ast Z)F \right)\,.
\eea
The boundary terms are given by
\bea
{\cal L}_{(1,1) \, bound} &=&   2\kappa v_nv^n + 2\kappa  \chi_+ \c^m\p_m \chi_+ + 4\kappa  \chi_+\lambda_- + 2\kappa   \psi_- \c^m\p_m \psi_-\\
 && - 2\kappa (f+\p_3a)^2 + \kappa  bb\\
 && \frac{1}{2} \xi^\ast\c^3 \xi + \frac{1}{2} F^\ast Z + \frac{1}{2}  Z^\ast F + \frac{1}{2} \p_3(Z^\ast Z)
 \eea

 If we go on-shell, we could perform a naive EL variation in $F$ we come to the conclusion that $Z$ must be zero on the boundary. Alternatively we might choose $F=0$ as a boundary condition upfront, but after looking at the bulk equation for $F$ we come to the same conclusion that $Z=0$ as a boundary condition.  More sophisticated would be to look at the boundary  ${ \cal N} =(1,1)$ multiplets and see that $\p_3 Z + F$ is an appropriate boundary condition.  Since this choice eliminates $F$ from the boundary action, we may simply integrate out the $F$ term and generate the bosonic potential for on-shell fields.
 Also following the discussion in the section 7 (i.e. pure chern-simons) it is natural to choose gauge sector boundary conditions in which $b= f+\p_3a =0$.
 In this case, the lagrange multiplier which enforces  $b = \frac{1}{4\kappa}Z^\ast Z$ means that the $b=0$ boundary condition is equivalent to fixing $Z=F=0$ on the boundary.  This is a little disheartening since the restrictions appear so strong.  However, we shall see that when we have a superpotential that this result changes.

\section{ABJM with Boundary}

\subsection{ABJM review}
With a single gauge group factor the maximal extension of Chern-Simons matter theories seems to be ${\cal N}=3$ \cite{Kao:1993gs}.  From an ${\cal N}=2$ superspace perspective this theory is built from pairs of chiral matter fields transforming in conjugate representations of the gauge group and a particular superpotential \cite{Gaiotto:2007qi}. In component form this can be recast in a way with manifest $SU(2)_R$ symmetry.

The ABJM model generalizes the ${\cal N}=3$ theory by having two gauge group factors and and two conjugate sets of bi-fundamental matter.  We summarize this action in ${\cal N}=2$ superspace as formulated in \cite{Benna:2008zy}.

The gauge fields are contained in two $U(N)$ adjoint superfields ($V$ and $\hat{V}$). We will suppress all gauge indices. The kinetic terms for these gauge fields are Chern-Simons but at opposite levels $k$ and $-k$.  In superspace the non-abelian Chern-Simons action is a little complicated and is given by
\bea
S^{CS}[V]= \kappa \int d^3 x \int d^4\th \int_0^1 dt \textrm{Tr} V \bar{D}^\a (e^{tV} D_\a e^{-tV})\, .
\eea

 The matter is described by bi-fundamental chiral superfields $\Z^{A}$ and $\W_A$ transforming respectively in the $(N,\bar{N})$ and $(\bar{N},N)$ of the group. The global flavor index takes values $A = \{1,2\}$. The components of $\Z^A$ ($\W_A$) are a complex scalar, $Z^A$ ($W_A$), a fermion, $\xi_\a^A$ ($\omega_{A\a}$), and an auxiliary complex scalar $F^A$ ($G_A$).  The four complex scalars  $Z^A$ and $W_A$ encode the transverse position of the membrane:
 \bea
 Z_1 = X^1 + iX^5 & W^1 = X^{3\dag} + iX^{7\dag}\\
 Z_2 = X^2 + iX^6 & W^1 = X^{4\dag} + iX^{8\dag}\, .
 \eea
 Making this split of the transverse scalars means giving up on having manifest $SO(8)$ R-symmetry.

  The kinetic terms for the matter fields are, in our conventions,
 \bea
S^{Mat}&=& \int d^3 x \int d^4\th \textrm{Tr} \left( \bar{\Z}_A e^{2V}\Z^Ae^{-2\hat{V}}  +  \bar{\W}^A e^{2\hat{V}}\W_Ae^{-2V} \right)\
\eea
  The superpotential is given by
\bea
S^{pot}&=&  \int d^3x  \int d^2\th W(\Z, \W)  + \int d^3x  \int d^2\bar{\th} W(\bar{\Z}, \bar{\W})\, ,
\eea
where
\bea
W = \frac{1}{\kappa} \e_{AC}\e^{BD} \textrm{Tr}\left(\Z^A \W_B\Z^C \W_D  \right) \, , &  \bar{W} = \frac{1}{\kappa} \e^{AC}\e_{BD} \textrm{Tr}\left(\bar{\Z}_A \bar{\W}^B\bar{\Z}_C \bar{\W}^D  \right)\,.
\eea
 This superpotential has a manifest $SU(2)\times SU(2)$ global symmetry and since we are working in ${\cal N}=2$ superspace there is also a $U(1)_R$.  In fact, with the correct normalisation for the superpotential, which depends on the Chern-Simons level, the theory enjoys an enhanced $SU(4)_R$ symmetry \cite{Benna:2008zy}.  The ABJM model is therefore ${\cal N} = 6$ supersymmetric.  The geometric reason for this $\frac{3}{4}$ maximal supersymmetry is that the transverse scalars actually describe a $\mathbb{Z}_k$ orbifold of $\mathbb{C}_4$.   For $k=1,2$ this quotient should preserve all the supersymmetry \cite{Aharony:2008ug}, however the details of this enhancement are subtle.

 In summary the full ABJM model is then given by
\bea
kS^{CS}[V] - kS^{CS}[\hat{V}] + S^{Mat} + S^{Pot}\, .
\eea
The full bulk action can be found in \cite{Benna:2008zy} and is characterized by a quartic bose-fermi interaction and sextic bosonic potential.

 \subsection{$U(1)\times U(1)$ ABJM with boundary}

Given the complexity of the non-abelian Chern-Simons term it is natural to start with the most basic $U(1)\times U(1)$ theory. In this case the superpotential obviously vanishes.  Also, because the fields commute, it turns out that all of the matter interactions disappear once auxiliary fields are integrated out.  Without boundary the theory is simple and free;
\bea
S^{bulk}_{U(1)\times U(1)} &=& \int d^3 x \, 4\kappa \epsilon^{\mu\nu\rho}\left(v_\mu\p_\nu v_\rho - \hat{v}_\mu\p_\nu \hat{v}_\rho \right)\\
 && \,\,\,  +F^AF^\ast_A - {\cal D}^\mu Z^\ast_A {\cal D}_\mu Z^A  - \xi_A^\ast \c^\mu {\cal D}_\mu \xi^A\\
 && \,\,\, + G_AG^{A\ast}  - {\cal D}^\mu W^{A\ast} {\cal D}_\mu W_A  - \omega^{A\ast} \c^\mu {\cal D}_\mu \omega_A\, ,
 \eea
where the covariant derivative acts as  ${\cal D}_\mu Z  =  \p_\mu Z + i v_\mu Z - i Z \hat{v}_\mu$ and with opposite charges on $W$.  In the above expression we have eliminate all auxiliary fields except $F$ and $G$ for reasons that will become clear shortly.

 With a boundary we can essentially read off the additional terms me must add to restore half the supersymmetry from section 7.

 In the ${\cal N}=(2,0)$ procedure we find we must include the following boundary terms:
 \bea
 {\cal L}^{(2,0) \, bound}_{U(1)\times U(1)} &=&  2k \left(  v_nv^n + \chi_- \c^m\p_m \chi_- +  \psi_- \c^m\p_m \psi_- + \p_m a\p^m a\right) \\
 && - 2k\left(  \hat{v}_n\hat{v}^n + \hat{\chi}_- \c^m\p_m  \hat{\chi}_- +  \hat{\psi}_- \c^m\p_m \hat{\psi}_- + \p_m\hat{a}\p^m \hat{a}\right)      \\
&& \,\, +\xi^\ast_{A-} \xi^A_+   +\omega^\ast_{A-} \omega^A
 \eea
 In this expression,  we have used the the Lagrange multiplier equation for $g$ and $\hat{g}$ to give values to the auxiliary fields $b$ and $\hat{b}$. These provide canceling contributions in the abelian case.  Since we have chiral ${\cal N}=(2,0)$ supersymmetry, it comes as no surprise that  we have a chiral action with propagating fermions of the same chirality.

In the  ${\cal N}=(1,1)$ case
\bea
\label{ABJM11boundarytheory}
 \nonumber {\cal L}^{(1,1) \, bound}_{U(1)\times U(1)} &=& 2k \left( v_nv^n + 2\chi_+\lambda_- +  \chi_+ \c^m\p_m \chi_+ +  \psi_- \c^m\p_m \psi_- - (f+\p_3a)^2 + \frac{1}{2}bb \right) \\
  \nonumber && -2k \left( \hat{v}_n\hat{v}^n + 2\hat{\chi}_+\hat{\lambda}_- +\hat{\chi}_+ \c^m\p_m \hat{\chi}_+ +  \hat{\psi}_- \c^m\p_m \hat{\psi}_-
   - (\hat{f}+\p_3\hat{a})^2 + \frac{1}{2}\hat{b}{b} \right)\\
  \nonumber &&   +\frac{1}{2}\xi_A^\ast\c^3 \xi^A +\frac{1}{2}F_A^\ast Z^A +  \frac{1}{2}Z_A^\ast F^A + \frac{1}{2}\p_3(Z_A^\ast Z^A)\\
  && +\frac{1}{2}\omega^{A\ast}\c^3 \omega_A +\frac{1}{2}G^{A\ast} W_A +  \frac{1}{2}W^{A\ast} G_A + \frac{1}{2}\p_3(W^{A\ast} W_A)
\eea
Here we see that $G_A$ and $F^A$ occur as boundary couplings.

We have preserved $(1,1)$ supersymmetry in the parity invariant ABJM model. This  strongly suggests that boundary theory  should have equal number of left and right movers. It is not chirality invariant because these are different fields.

To gain an immediate physical understanding we go on-shell. We pick boundary conditions that are consistent with the parity invariance of the ABJM model and the supersymmetry. The boundary (1,1) superfields detailed in the appendix readily tell us how to choose supersymmetric boundary conditions\footnote{we drop the tilde-hat notation of the appendix and understand that hatted quantities correspond to the hatted vector multiplet}.

For the vector multiplet $V$ we choose the following boundary condition:
\bea
0 = V_\a = \psi_- + \th_{1-}(f+\p_3 a) -\th_{2+}\c^1 v_- + \th_{2+}\th_{1-} (\lambda_- - \c^1 \p_-\chi_+)\,.
\eea
For the other vector multiplet $\hat{V}$ we choose:
\bea
0=\hat{U}_\a = \hat{\chi}_+ - \th_{1-}\c^1 \hat{v}_+ - \th_{2+} (\hat{b} - (\hat{f} + \p_3\hat{a}) + \th_{2+}\th_{1-} (\hat{\eta}_+ + \c^1 \p_+\hat{\psi}_-)\,.
\eea
Notice that we have $v_-=0$ and $\hat{v}_+=0$, which is compatible with parity. These conditions are exactly what we have seen is required to produce the combination of a chiral and anti-chiral FJ action in the pure gauge (no matter) theory.

The lowest component of these two boundary superfields show that $\psi_-$ and $\hat{\chi}_+$ are constrained to zero.  We may also set the gauginos appearing in the boundary conditions to zero, i.e. $\lambda_-=\hat{\eta}_+ =0$.

For the auxiliary scalars demanding $f+\p_3 a=\hat{b} - (\hat{f} + \p_3\hat{a})=0$ can only be compatible with parity if both $\hat{b}$ and $b$  are also set to zero on the boundary.

As we saw in the earlier example in section 8, the appropriate boundary conditions on the matter seem to be $F+\p_3Z = G+\p_3W =0$.  In the Abelian scenario this, as before, forces $Z=W=0$ on the boundary. This will not be true in the non-abelian case because there is a superpotential.

If we plug in the trivial i.e. algebraic and non-derivative boundary conditions in to the (1,1) boundary terms (\ref{ABJM11boundarytheory}) we are simply left with
\bea
  {\cal L}^{(1,1) \, bound}_{U(1)\times U(1)} &=& 2k \chi_+\c^m\p_m \chi_+  - 2k\hat{\psi}_- \c^m\p_m \hat{\psi}_-
  \eea
This boundary theory quite clearly has two propagating fermions of opposite chirality and is essentially non-chiral.

 \subsection{Towards $U(N)\times U(N)$ ABJM with boundary}
 As we have seen the non-abelian kinetic term is a very complicated affair.  In principle we could, by following the procedure of section 6, construct the boundary action to preserve half the supersymmetry. (For a treatment of ${\cal N}=1$ non-abelian super Chern-Simons using similar techniques to us see \cite{Sakai:1989nh}).   However, if one wanted to look at this in component form it would take significant effort.   There are also addition complications concerning field redefinitions and gauge fixing.  Furthermore we would have to establish the correct additional terms required to remove the gaugino boundary interactions.

  In this paper we don't intend to complete all of the above.  Instead we will make a couple of sensible assumptions that will allow us to learn about the matter sector.

  We assume that if, and only if, we were able to eliminate an auxiliary field boundary interaction in the abelian case through the addition of separate boundary actions we will be able to do so for the non-abelian case.  The only difference will be the obvious inclusion of a trace.

    Although we will only have partial knowledge of the gauge sector boundary terms (e.g. we do not know any commutator terms) we will have full knowledge of the matter sector.  This will be enough to inform us about a boundary potential for the bosonic matter fields.

\subsection{${\cal N} = (2,0)$ supersymmetry }
We first consider the case where we preserve manifest ${\cal N} = (2,0)$ supersymmetry.

In what follows we shall turn our attention to the just the boundary contributions for the bosonic matter fields.  We find the following boundary terms
\bea
{\cal L}_{bound}^{(2,0)} = -\kappa b^ab^a + \frac{1}{2}b^a \textrm{Tr}\left( T^a(ZZ^\dag  - W^\dag W)\right)  +\kappa\hat{b}^a\hat{b}^a  -  \frac{1}{2}\hat{b}^a \textrm{Tr} \left(T^a(Z^\dag Z  - W W^\dag)\right) + \dots
\eea
where the dots indicate boundary contributions from fermions in the matter multiplet and terms generated by the gauge multiplet which don't interact with the matter. The abelian contribution to these omitted terms can be read off from the constructions in the preceding section.  Because we are preserving  ${\cal N} = (2,0)$ supersymmetry there is no boundary contribution from the superpotential.

In the bulk, we find that $g$ and $\hat{g}$ are Lagrange multipliers enforcing $b$ and $\hat{b}$ to take a particular value given by
\bea
 b^a = \frac{1}{4\kappa} \textrm{Tr}\big( T^a(ZZ^\dag - W^\dag W) \big) \, , &  \hat{b}^a = \frac{1}{4\kappa} \textrm{Tr}\big( T^a(Z^\dag Z- W W^\dag ) \big)\, .
\eea
We may make the above replacement into the boundary terms and find
\bea
{\cal L}_{bound}^{(2,0)} = -\frac{1}{16\kappa}\textrm{Tr}[(Z^\dag Z- WW^\dag)^2 - (ZZ^\dag - W^\dag W   )^2 ]  + \dots\,.
\eea
 A key observation is that we now have a quartic boundary scalar potential. We interpret this as being the effect of a five-brane.
 The consequence of this boundary potential can be seen in the field equations and Euler-Lagrange boundary conditions.  The matter field $Z$ obeys a natural boundary condition of the form
\bea
\p_3 Z^A + \frac{1}{8\kappa}[Z^A(Z^\dag Z- WW^\dag) - (ZZ^\dag - W^\dag W   )Z^A ] = 0\,.
\eea
It is helpful to introduce a three-bracket given by
\bea
[A, B; C] = AC^\dag B- BC^\dag A
\eea
in order to make contact with the Bagger-Lambert formulation of the ABJM model\cite{Bagger:2008se}. One can re-write the boundary condition using this bracket as
\bea
\p_3 Z^A- \frac{1}{8\kappa}( [Z^B, Z^A; Z_B^\dag] + [Z^A, W^{\dag B}; W_B] )=0\,.
\eea
This equation (together with the similar contribution for $W$) can be seen in the \textquoteleft D-term\textquoteright\,  BPS equation found in \cite{Hanaki:2008cu} by looking at the Hamiltonian of the ABJM model.  However, the full corresponding BPS equations also include a constraint $\e_{AC}\e^{BD} W_BZ^CW_D=0$ which we have not observed.   When only half the scalars, e.g the $Z^A$, are excited this constraint is solved and the remaining BPS equation simplifies to
\bea
\label{fuzzyeqn}
\p_3 Z^A - \frac{1}{8\kappa}[Z^B, Z^A; Z_B^\dag]=0\, .
\eea
 This equation should yield fuzzy funnel solutions describing the membrane ending on the five-brane.  However, the symmetry of this equation is only $SU(2)\times U(1)$ whereas the Basu-Harvey equation describing fuzzy three-spheres has $SO(4)$ symmetry.  Solutions of (\ref{fuzzyeqn}) have been found and are thought to represent fuzzy $S^3/\mathbb{Z}_k$ \cite{Hanaki:2008cu}.  In \cite{Nastase:2009ny} fluctuations of the fuzzy funnel were analyzed in a large $k$ limit where a perturbation theory can be used. This indicated an underlying fuzzy $S^2$ structure rather than the perhaps expected fuzzy $S^3$.

\subsection{${\cal N} = (1,1)$ supersymmetry}
We turn to ${\cal N} = (1,1)$ case. In this case we find the following contributions to the bosonic boundary term
\bea
{\cal L}_{bound}^{(1,1)}&=&\frac{1}{2}Z^\dag(F+\p_3Z) + \frac{1}{2}W(W^\dag + \p_3 G^\dag  + h.c.\\
&& -\frac{1}{2} b(ZZ^\dag - W^\dag W) + \frac{1}{2} \hat{b}(Z^\dag Z - W W^\dag) - \kappa bb + \kappa \hat{b}\hat{b}\\
&& -\frac{1}{8\kappa} \epsilon_{AC}\epsilon^{BD} Z^AW_BZ^CW_D + h.c. + \dots \, ,
\eea
where again the dots indicate the fermions and the decoupled gauge sector.

As with the abelian scenario the presence of auxiliary fields in this action makes it hard to understand the on-shell nature of the theory.  If we choose the same boundary conditions as the abelian case i.e.
\bea
0 = b = \hat{b}=F^A+\p_3Z^A = G_A+\p_3 W_A
\eea
then the boundary action reduces to
\bea
\label{remainingboundaryterm}
{\cal L}_{bound}^{(1,1)}&=& -\frac{1}{8\kappa} \epsilon_{AC}\epsilon^{BD} Z^AW_BZ^CW_D + h.c. + \dots \, .
\eea
Unlike the abelian case however, the $b$ boundary condition does not require that $Z=W=0$. It does constrain the matter fields to obey
\bea
\label{11BPSconstraint}
 b^a = \frac{1}{4\kappa} \textrm{Tr}\big( T^a(ZZ^\dag - W^\dag W) \big) = 0  \, , &  \hat{b}^a = \frac{1}{4\kappa} \textrm{Tr}\big( T^a(Z^\dag Z- W W^\dag ) \big) = 0
\eea
on the boundary.
We may also make use of the bulk equation for $F$ and $G$ together with the $b$ boundary condition to write the matter boundary condition as
\bea
\label{11BPSequations}
\p_3 Z^A - \frac{1}{4\kappa} \epsilon^{AC}\epsilon_{BD} W^{\dag B} Z^\dag_CW^{\dag D} = 0\,,  &\,
\p_3 W_A + \frac{1}{4\kappa} \epsilon_{AC}\epsilon^{BD} Z^{\dag }_BW^{\dag C} Z^{\dag }_D= 0
\eea
These equations can be recognised as the \textquoteleft F-term\textquoteright\, BPS equations found in \cite{Hanaki:2008cu} by the Bogomolny completion of the Hamiltonian.   The constraints (\ref{11BPSconstraint}) also imply the constraints found by the Bogomolny trick, although here they are a little stronger.

Note that after invoking these boundary conditions there remains a quartic boundary potential, which is not set to zero, and is given by (\ref{remainingboundaryterm}).   Upon performing an EL variation of the bulk+boundary action the total derivative picked up from varying the scalar kinetic terms combines with the variation of the boundary potential to reproduce the boundary conditions.

\section{Discussion}

This paper is the first step towards the study of open interacting
membranes. The ultimate aim is to gain insight into the fivebrane as
a theory of open membranes though as yet we are still far from that
goal. In spite of this, the reproduction of the BPS equations as supersymmetric
boundary equations encourages us that we are on the right path to
understanding more about the interacting self-dual string.

(As an aside, the gauge sector of the theory is interesting in its own
right as its boundary theory produces an interesting WZW model. The
role of parity and the resulting chirality in the WZW model is
particularly interesting).

There is still a great deal to understand. In particular one would
like to see the role of the quartic potential on the boundary from the
fivebrane perspective. A more direct question is to understand
the boundary equation arising from the (1,1) supersymmetry. As stated
this has been observed before as a BPS equation but so far the
solutions are not known and the brane interpretation is open. Solving
this equation and interpreting the solutions would hopefully provide
some insight. The relation between the two choices of supersymmetry are
also interesting and perhaps there is a map between solutions.

The most important limitation to this work was that using
the superspace method described above meant that it was easiest to deal
with only a manifest ${\cal N}=2$ supersymmetry. Obviously
extending these results to higher supersymmetry would be of great
interest though somewhat technically demanding. Essentially, one would
like a complete classification of the boundaries preserving different
amounts of supersymmetry, beginning with membranes that also preserve
differing amounts of supersymmetry. We also did not carry out a
rigorous derivation using the full non-abelian superspace action;
although we expect no surprises it would be good to have this further developed.

There is a very interesting complimentary approach to this work
that we have as yet not explored. To study open interacting membranes,
one could use an ABJM style
brane set up in IIB with the addition of a suitable boundary and thus
the inclusion of an additional NS5 brane. One could then be able to
utilise the work of Gaiotto and Witten \cite{Gaiotto:2008sa}
 in this ABJM configuration to learn about the open membrane. It would
 be interesting to see from this perspective whether the different choices
of supersymmetry preserved by the boundary are related to symmetries
in the branes set up. (The changing of the type of supersymmetry seems
reminiscent of T-duality).

\section{Acknowledgements}

We wish to acknowledge helpful discussions with James Bedford, Oren
Bergman, Sergey Cherkis, Neil Lambert, Andrew Low, George Papadopoulos, Constantinos
Papageorgakis and Sanjaye Ramgoolam. We also wish to thank Will Black
for help with the tex file and illustrations.
DCT acknowledges the support of an STFC doctoral grant. DSB
acknowledges the support of an STFC rolling grant and thanks DAMTP for
hospitality during completion of this work.

\section{Appendix 1: ${ \cal N}= 1$ Supersymmetry Conventions}
We broadly follow Superspace. Index contraction and manipulation is given by
\bea
\th^\a = C^{\a \b} \th_\b \, , & \th_\b = \th^\a C_{\a \b} \, ,  &\th_\a \th_\b = -C_{\a \b} \th^2 = -\frac{1}{2} C_{\a \b} \th^\c \th_\c \, ,
\eea
where
\bea
C_{\a \b} = - C_{\b \a} = -C^{\a \b} = \left( \begin{array}{cc} 0 &-i \\ i & 0 \end{array} \right) \, , &
C_{\a\b} C^{\c\d} = \d_{[\a}^\c \d_{\b]}^\d \, .
\eea
When spinor indices are suppressed index contraction is always top right to bottom left $(\searrow)$. With this convention it is unnecessary to show conjugation with an overbar.
Gamma matrices obey the Clifford algebra
\bea
\{\c^\mu , \c^\nu \}_{\a \b}= 2g^{\mu \nu}C_{\a \b}
\eea
where
 \bea
\gamma^\mu_{\a \b} \equiv \gamma_\a^{\mu \c} C_{\c \b} = \gamma^\mu_{\b \a}\, ,  &
\c^\mu \c^\nu = \eta^{\mu \nu} + \e^{\mu \nu \r} \gamma_\r \, .
\eea
Differentiation and integration is summarized by
\bea
\p_\a \th_\b = C_{\a \b} \, ,&    \int d^2 \theta \theta^2 = -1 \,.
\eea
The SUSY charge and covariant derivative are
\bea
Q_\a = \p_\a - (\c^\mu \th)_\a \p_\mu\, , \\
D_\a = \p_\a + (\c^\mu \th)_\a \p_\mu \, ,
\eea
and the algebra is
\bea
\{Q_\a, Q_\b\} = - \{D_\a D_\b\} = 2\c^\mu_{\a \b} \p_\mu\, .
\eea
The covariant derivatives satisfy the following identities
\bea
D_\a D_\b &=& - \c^\mu_{\a \b} \p_\mu - C_{\a \b} D^2 \, ,\\
D^2 D_\a &=& - D_\a D^2 = (\c^\mu D)_\a \p_\mu\, ,\\
(D^2)^2 &=& \Box \, ,\\
D^\a D_\b D_\a &=& 0 \,.
\eea
A scalar superfield is given by
\bea
\Phi = a + \th \psi - \th^2 f  = (a, \psi , f)\, ,
\eea
and a spinor superfield by
\be
\Gamma_\a = \chi_\a - \th_\a M + (\c^\mu \th)_\a v_\mu - \th^2[ \lambda_\a + (\c^\mu \p_\mu \chi)_\a]  = (\chi, M, v_\mu , \lambda) \, .
\ee
We use early Greek letters to denote spinor indices, late Greek for 3-dimensional space-time indices (with $x_\mu = (x_0, x_1, x_3)$) and Latin indices for two-dimensional space-time ($x_m=  (x_0, x_1)$).   We assume Lorentzian  (-++) signature and $\epsilon^{013} = +1$.

\section{Appendix 2:  ${ \cal N}= 2$ Supersymmetry Conventions}
In three-dimensions ${\cal N}=2$ superspace is realized by taking the Grassman coordinates to be complex.
For our purposes it is convenient to express these in terms of an ${\cal N}=1$ decomposition by writing
\bea
\th = \frac{1}{\sqrt{2}}(\th_1 + i \th_2)\, , & \bar{\th} = \frac{1}{\sqrt{2}}(\th_1 - i \th_2)\,
\eea
so that the  ${\cal N}=2$  superspace covariant derivatives
\bea
D_\a = \p_\a +(\c^\mu \bar{ \th})_\a \p_\mu\, , & \bar{D}_\a = \bar{\p}_\a +(\c^\mu  \th)_\a \p_\mu\,.
\eea
are decomposed as
\bea
D_\a = \frac{1}{\sqrt{2}}(D_1 - i D_2)\, , & \bar{D}_\a = \frac{1}{\sqrt{2}}(D_1 + iD_2)\,
\eea
with
\bea
D_i = \frac{\p}{\p \th_i} + (\c^\mu  \th_i)_\a \p_\mu \quad i = \{1,2\} \, ,
\eea
satisfying the algebra
\bea
\{ D_{i\a}\,, D_{j\b} \} = -2 \d_{ij} \c^\mu_{\a\b} \p_\mu\, .
\eea
The  ${\cal N}=2$ chiral superfield  obeys
\bea
\bar{D} \Phi = 0  \Leftrightarrow D_2 \Phi = i D_1 \Phi
\eea
and can be expressed as
\bea
\Phi = X(\th_1) + i\th_2 D_1 X + D_1^2 X \th^2_2
\eea
where $X(\th_1)= a+\th_1 \psi - f \th_1^2$ is a complex ${\cal N}=1$  superfield.
The ${\cal N}=2$  vector field is a real superfield obeying
\bea
V = A(\th_1) + \th_2 \Gamma - \th_2^2 (B - D_1^2A)
\eea
with $A$ and $B$ real ${\cal N}=1$  scalar superfields and $\Gamma_\a$ a real ${\cal N}=1$  spinor superfield.  The ${\cal N}=2$  gauge transformations are
\bea
\d_{gauge} V = i[ \Phi-\bar{\Phi}]  \Rightarrow \left\{ \begin{array}{l}
\d_{gauge} A = i[X-\bar{X}]\\
\d_{gauge} \Gamma_\a = - D_{1\a} (X+ \bar{X}) \\
\d_{gauge} B = 0 \end{array}\right\}\, .
\eea
The arbitrary shift in $A$ is usually used to gauge fix this field to zero.

\section{Appendix 3: Multiplet Decomposition}
\subsection{${\cal N} =1 \rightarrow {\cal N} =(1,0)$ }
There is a simple procedure to obtain half supersymmetric multiplets from a fully supersymmetric one.
We introduce projectors
\be
(P_\pm) = \frac{1}{2} (C_{\a\b} \pm \gamma^3_{\a\b})
\ee
and write the scalar multiplet as
\bea
A &=& a + \theta \psi - \theta^2 f\\
&=& \exp ( - \th_+\th_- \p_3 ) \left( \hat{A}(\th_+) + \th_-^\a \hat{A}_\a (\th_+) \right)\\
&=& \exp (+ \th_+\th_- \p_3 )\left( \tilde{A}(\th_-) + \th_+^\a \tilde{A}_\a (\th_-) \right)
\eea
where hatted objects are now 1+1 dimension superfields whose supersymmetry is generated by $\e_+ Q'_- \equiv \e_+[ \frac{\p}{\p \th_+} -  \c^m\th_+ \p_m]$ and tilded objects are  1+1 dimension superfields whose supersymmetry is generated by $\e_- Q'_+ \equiv \e_-[ \frac{\p}{\p \th_-} -  \c^m\th_- \p_m]$.  A useful relation is $\frac{\p}{\p \th^\a_+}  \th_+^\b = P_{-\a}^{\phantom{-\a} \b}$.

One can take the spinor multiplet project with $P_\pm$ and perform a similar decomposition
\bea
\Gamma^\pm_{\a} &=& \exp ( - \th_+\th_- \p_3 ) \left( \hat{\Gamma}^\pm_\a (\th_+) + \th_-^\b \hat{\Gamma}^\pm_{ \b \a} (\th_+) \right)\\
&=& \exp (+ \th_+\th_- \p_3 )\left( \tilde{\Gamma}^\pm_\a (\th_-) + \th_+^\b \tilde{\Gamma}^\pm_{ \b \a} (\th_-) \right)\,.
\eea
In principle the $\hat{\Gamma}^\pm_{ \b \a}$ and $\tilde{\Gamma}^\pm_{ \b \a}$  are Lorentz reducible multiplets so we must take symmetric and anti-symmetric parts to discover the correct irreducible superfields. However one find that these fields are either scalar or vector like and not a combination of both. To be explicit we can write
\bea
\th_-^\b \hat{\Gamma}^+_{ \b \a}  = -\th^{\b}_-\c_{\b\a}^m \hat{\Sigma}^+_m\,, & \th_-^\b \hat{\Gamma}^-_{ \b \a}  = -\th_{-\a} \hat{\Sigma}^- \,, \\
\th_+^\b \tilde{\Gamma}^+_{ \b \a}  = -\th_{+\a} \tilde{\Sigma}^+\,, & \th_+^\b \tilde{\Gamma}^-_{ \b \a}  = -\th^{\b}_{+}\c_{\b\a}^m \tilde{\Sigma}^-_m\, .
\eea
In components we have
\bea
 \hat{A}&=& a+ \th_+ \psi_- \\
 \hat{A}_\a &=& \psi_+ + \th_+ \left( - f + \p_3 a\right)\\
 \hat{\Gamma}^+_\a&=& \chi_+ +\th_+\left(v_3-M\right)\\
 \hat{\Sigma}^+_m&=& v_m +  \th_+\left(\frac{1}{2}\c_m\lambda_+ + \p_m \chi_-\right)\\
 \hat{\Gamma}^-_\a&=& \chi_- - \th_+\c^m v_m \\
  \hat{\Sigma}^- &=& M+v_3 - \th_+\left(\lambda_- - 2 \p_3 \chi_- + \c^m \p_m \chi_+\right)\\
 \nonumber \,&&\\
  \tilde{A}&=& a+\th_-\psi_+\\
 \tilde{A}_\a&=& \psi_- - \th_- \left(f+\p_3 a\right)\\
  \tilde{\Gamma}^+_\a&=&   \chi_+ - \th_- \c^m v_m\\
\tilde{\Sigma}^+ &=&M-v_3 - \th_-\left(\lambda_+ +2 \p_3 \chi_+ + \c^m\p_m \chi_-\right)\\
 \tilde{\Gamma}^-_\a&=& \chi_- - \th_- \left(M+v_3\right)\\
 \tilde{\Sigma}^-_m &=&  v_m + \th_- \left(\frac{1}{2} \c_m \lambda_- + \p_m \chi_+\right)
\eea
To get to these forms of the superfields one actually must do a little work and use identities like
\bea
(\th_-\c^m)_\a \th_+ \c_m \c_n \p^n \chi_- = 2 (\th_-\c^m)_\a \th_+ \p_m \chi_-\,.
\eea
Also when calculating the correct boundary conditions one should bear in mind identities like, for example,
\bea
(\th_- \c^m)_\a \hat{\Sigma}_m^+ = (\th_- \c^1)_\a \left( \hat{\Sigma}_0^+ + \hat{\Sigma}_1^+  \right)=  (\th_- \c^1)_\a \hat{\Sigma}_{m=+}^+  \,.
\eea
\subsection{${\cal N} =2 \rightarrow {\cal N} = (2,0)  \, \textrm{or}\,  {\cal N} = (1,1)$ }
Starting with the most general superfield
\bea
V= A(\th_1) + \th_2 \Gamma(\th_1) - \th_2^2 C(\th_1)
\eea
we decompose this into 2d ${\cal N} = (1,1)$ or ${\cal N} = (2,0)$ multiplets given by
\bea
V&=& \exp ( - \th_{2+}\th_{2-} \p_3 )\exp ( + \th_{1+}\th_{1-} \p_3 )\left( \tilde{\hat{V}}(\th_{1-}, \th_{2+}) + \th_{1+}^\a\tilde{\hat{V}}_\a +\th_{2-}^\a \tilde{\hat{U}}_\a + \th_{2-}\th_{1+}\tilde{\hat{U}}  \right)\\
&=&\exp ( - \th_{2+}\th_{2-} \p_3 )\exp ( - \th_{1+}\th_{1-} \p_3 )\left( \hat{\hat{V}}(\th_{1+}, \th_{2+}) + \th_{1-}^\a\hat{\hat{V}}_\a +\th_{2-}^\a \hat{\hat{U}}_\a + \th_{2-}\c^m\th_{1-}\hat{\hat{V}}_{m}   \right)
\eea
We can, in turn, express the components of these expansions in terms of the ${\cal N} = 1$ decompositions defined previously. We have for ${\cal N} = (1,1)$
\bea
\tilde{\hat{V}} &=& \tilde{A}\left(\th_{1-}\right) + \th_{2+}^\a \tilde{\Gamma}^{-}_\a\\
\tilde{\hat{V}}_\a &=& \tilde{A}_\a- (\c^m\th_{2+})_a\tilde{\Sigma}_m^-\\
\tilde{\hat{U}}_\a &=&  \tilde{\Gamma}^{+}_\a + \th_{2+\a} \left(-\tilde{C} + \p_3 \tilde{A}\right)\\
\tilde{\hat{U}}&=& -\tilde{\Sigma}^+  + \th_{2+}^\b \left( \tilde{C}_\b - \p_3\tilde{A}_\b\right)
\eea
and for ${\cal N} = (2,0)$:
\bea
\hat{\hat{V}} &=& \hat{A}\left(\th_{1-}\right) + \th_{2+}^\a \hat{\Gamma}^{-}_\a\\
\hat{\hat{V}}_\a &=& \hat{A}_\a- \th_{2+\a} \hat{\Sigma}^- \\
\hat{\hat{U}}_\a &=&  \hat{\Gamma}^{+}_\a + \th_{2+\a} \left(-\hat{C} + \p_3 \hat{A}\right)\\
\hat{\hat{V}}_m &=&  \hat{\Sigma}^+_m - \frac{1}{2} \left(\th_{2+} \c_m\right)^\a \left(\hat{C}_\a - \p_3\hat{A}_\a\right)
\eea

For the case of the vector field
\bea
V(\th_1, \th_2) = A(\th_1) + \th_2 \Gamma (\th_1) - \th_2^2 (B(\th_1) - D_1^2A)
\eea
with components
\bea
A=(a, \psi, f) \, ,& B=(b, \eta, g) \, , & \Gamma = (\chi, M, v, \lambda)\, .
\eea
the half supersymmetric multiplets are
\bea
\hat{\hat{V}} &=& a + \th_{1+}\psi_-+\th_{2+}\chi_-+\th_{2+} \c^m\th_{1+}v_m\\
\hat{\hat{V}}_\a &=& \psi_{+\a} + \th_{1+\a}(-f+\p_3a)-\th_{2+\a}(M+v_3) +\th_{2+\a} \th_{1+}^\b(\lambda_- - 2\p_3\chi_- + \c^m\p_m \chi_+ )_\b \\
\hat{\hat{U}}_\a &=&   \chi_{+\a} + \th_{1+\a}(v_3 -M) + \th_{2+\a}(f+\p_3a -b) - \th_{2+} \c_m\th_{1+} \c^{m\,\b}_{\a} (\eta_- +\c_m\p_m \psi_+ -2 \p_3\psi_-)_\b  \\
\nonumber \hat{\hat{V}}_m &=& v_m  + \th_{1+}(\frac{1}{2} \c_m \lambda_+ + \p_m \chi_- ) - \th_{2+}(\frac{1}{2} \c_m \eta_+ + \p_m \psi_- )\\
 && \quad\quad- \th_{2+} \c_m\th_{1+} (\frac{1}{2}(-g + \p_3b + \p^m\p_ma))\\
\nonumber\\
\tilde{\hat{V}} &=& a + \th_{1-}\psi_++\th_{2+}\chi_- - \th_{2+}\th_{1-}(M+v_3)\\
\tilde{\hat{V}}_\a &=& \psi_{-\a} - \th_{1-\a}(f+\p_3a) +(\th_{2+}\c^m)_\a v_m+ \th_{2+}\th_{1-}(\lambda_- +\c^m\p_m \chi_+)\\
\tilde{\hat{U}}_\a &=& \chi_{+\a} - (\th_{1-}\c^m)_\a v_m   - \th_{2+\a} (b-(f+\p_3a))+ \th_{2+}\th_{1-}(\eta_+ +\c^m\p_m \psi_-)\\
\nonumber \tilde{\hat{U}}&=& v_3-M +\th_{1-}(\lambda_+ +2\p_3 \chi_+ + \c^m\p_m \chi_-) + \th_{2+}(\eta_- + \c^m\p_m \psi_+ - 2 \p_3 \psi_-)\\
 && \quad\quad- \th_{2+}\th_{1-} (g- \p_m\p^ma - 2 \p_3\p_3 a - 2 \p_3 f - \p_3b)
\eea

\bibliographystyle{alpha}

\end{document}